\documentclass[12pt,thmsa]{article}
\usepackage{sw20lart}

\input{tcilatex}
\begin{document}

\author{Emilio Santos \and Departamento de F\'{i}sica. Universidad de Cantabria.
Santander. Spain}
\title{On a heuristic point of view concerning the motion of matter. From random
metric to Schr\"{o}dinger equation.}
\date{}
\maketitle

\begin{abstract}
The motion of a particle is studied in a random space-time metric, using a
non-relativistic approximation. The randomness induces a diffusion of the
particle in coordinate space. Hence it is shown that the evolution of the
probability density of the particle\'{}s positions is given by
Schr\"{o}dinger equation.

\end{abstract}

\section{\protect\smallskip Introduction}

Einstein never accepted quantum mechanics as a fundamental theory of nature.
Furthermore, he believed that a fundamental theory could not be found
starting from the current form of quantum mechanics (e.g. by adding hidden
variables) but within a completely different framework, probably that of
general relativity. In his own words: ``I do not believe that quantum
mechanics will be the starting point in the search for this basis just as
one cannot arrive at the foundations of mechanics from thermodynamics or
statistical mechanics''\cite{Pais}. The present paper attempts to explore
the possible derivation of a fundamental theory of motion in agreement with
Einstein\'{}s expectations. Here I shall restrict the study to motion with
small velocity in order to obtain a non-relativistic approximation. As we
shall see, although the starting point is different, the theory has some
similarity with the de Broglie-Bohm hidden-variables theory (or Bohmian
mechanics)\cite{Holland}. In particular, both theories assume the existence
of trajectories for the particles. The difference is that Bohmian mechanics
rests upon the \textit{hydrodynamic }interpretation of Schr\"{o}dinger
equation, in which particle trajectories never cross each other. In
contrast, in this paper I consider that the particle\'{}s motion consists of
a random motion superimposed to a smooth one. Thus we might speak about an 
\textit{aerodynamic} interpretation, which makes the approach similar to
stochastic mechanics\cite{Nelson}.

General relativity starts from the assumption that space-time may be curved.
The curvature can be derived from the space-time metric, once a coordinate
system is defined. The specific assumption in this paper is that the metric
is random. There are several reasons for this hypothesis. Firstly, noise is
quite natural in our very complex universe, therefore to assume the
existence of randomness is more plausible than to assume its absence.
Secondly quantum theory contains a random ingredient in the form of vacuum
fluctuations. At a difference with the standard assumption of
nineteenth-century physics that randomness is always associated to finite
(nonzero) temperature, quantum physics of the twentieth century contains
``zeropoint fields'', that is some randomness even at zero Kelvin. Thus I
propose that the motion of bodies should be always studied in a random
space-time metric. The randomness would be specified by defining the
probability distribution in the set of possible metrics, but in the present
paper I shall not state that distribution and use only some assumptions
about it.

\section{Motion in a random metric}

I consider a particle which is placed in \textbf{x}$_{1}$at time t$_{1}$and
in \textbf{x}$_{2}$ at time t$_{2}.$ It moves under the action of a
potential U$\left( \mathbf{x}\right) $ in a space-time metric 
\begin{equation}
ds_{\lambda
}^{2}=g_{00}c^{2}dt^{2}+2g_{0j}cdtdx^{j}+g_{jk}dx^{j}dx^{k},\;j,k=1,2,3,
\label{metric}
\end{equation}
where g$_{00},$g$_{0j}$ and g$_{ij}$ are functions of $\left( \lambda ;%
\mathbf{x,}t\right) $, $c$ being the velocity of light. As appropriate for
the non-relativistic approximation to be introduced later, we distinguish
the time, $t$, from the space coordinates, $\left\{
x^{1},x^{2},x^{3}\right\} ,$ these defining the position vector, $\mathbf{x}$%
. The randomness is taken into account assuming that there is a probability
density, $P\left( \lambda \right) ,$ $\lambda \in \Lambda ,$ in the set $%
\Lambda $ of space-time metrics. (For clarity in the physical arguments to
be introduced later, I use the notation $P\left( \lambda \right) d\lambda $
for the probability distribution, without any claim of mathematical rigour.)
Some ambiguity appears due to the freedom existing in general relativity for
the choice of the coordinate system, but this ambiguity should not produce
any confusion in what follows. In any case I shall assume that the choice of
coordinates for every metric is made so that the potential $U(\mathbf{x})$
has the same functional form in all of them.

For given $\lambda ,$ that is a fixed metric, the motion of the particle
would correspond to the minimum of the action 
\begin{equation}
A(\lambda )\equiv \int_{t_{1}}^{t_{2}}\left\{ -Mcds_{\lambda }-U\left( 
\mathbf{x}\right) dt\right\} =\int_{t_{1}}^{t_{2}}L\left( \lambda ;\mathbf{x,%
\dot{x},}t\right) dt=\min ,  \label{action}
\end{equation}
\begin{equation}
L\left( \lambda ;\mathbf{x,\dot{x},}t\right) \equiv -Mc\sqrt{%
g_{00}c^{2}+2g_{0j}c\dot{x}^{_{j}}+g_{ij}\dot{x}^{_{i}}\dot{x}^{_{j}}}%
-U\left( \mathbf{x}\right) ,  \label{lagrangean}
\end{equation}
where $\dot{x}^{j}$ are the velocities (time derivatives of the position
coordinates of the particle at a given time) and $U(\mathbf{x)}$ is the
potential. The generalization to include forces not derived from a potential
is straightforward, but it will not be considered in the present paper. An
alternative procedure to $\left( \ref{action}\right) $ would be to start
from the differential equation of motion for the particle in a given metric
(i. e. specified by a value of $\lambda )$ but a variational principle is a
more convenient starting point for my purposes. From now on I will speak
about the set of Lagrange functions $\left( \ref{lagrangean}\right) $,
rather than the set of metrics, but use the same label, $\Lambda ,$ for both
sets, which should not introduce any confusion.

In this paper I will consider only motions such that $\dot{x}^{_{j}}<<c,$
and consequently I shall use a non-relativistic approximation. Thus I may
replace $\left( \ref{lagrangean}\right) $ by an expansion to second order in
the velocities $\dot{x}^{j},$ which gives 
\begin{equation}
L\left( \lambda ;\mathbf{x,\dot{x},}t\right) =-M\left[ \sqrt{g_{00}}c^{2}+%
\frac{cg_{0j}}{\sqrt{g_{00}}}\dot{x}^{j}+\left( \frac{g_{jk}}{2\sqrt{g_{00}}}%
-\frac{g_{0j}g_{0k}}{2(\sqrt{g_{00}})^{3}}\right) \dot{x}^{j}\dot{x}%
^{k}\right] -U\left( \mathbf{x}\right) .  \label{lagr}
\end{equation}
For later convenience I define the ``mean Lagrange function $L_{0}$'', which
might be obtained by an average over $\lambda $ of $\left( \ref{lagr}\right) 
$, for fixed $\mathbf{x,}$ $\mathbf{\dot{x}}$ and $t$. It is 
\begin{equation}
L_{0}\left( \mathbf{x,\dot{x},}t\right) \equiv \int d\lambda P\left( \lambda
\right) L\left( \lambda ;\mathbf{x,\dot{x},}t\right) =-mc^{2}+\frac{1}{2}m%
\mathbf{\dot{x}}^{2}-U\left( \mathbf{x}\right) .  \label{lagr1}
\end{equation}
The simplicity of eq.$\left( \ref{lagr1}\right) $ is a consequence of the
invariance properties which I assume for the probability distribution of
metrics, that is invariance to translations, rotations, time-translations
and Galilean transformations. These invariance properties allow for the
renormalized mass $m$ to be different from the bare mass $M$. The Lagrange
function $\left( \ref{lagr}\right) $ may be written as a sum 
\begin{equation}
L\left( \lambda ;\mathbf{x,\dot{x},}t\right) =L_{0}\left( \mathbf{x,\dot{x},}%
t\right) +L_{1}\left( \lambda ;\mathbf{x,\dot{x},}t\right) ,  \label{lagr01}
\end{equation}
such that all randomness goes in the second term.

The probability distribution in the set of Lagrange functions induces a
probability distribution in the set of possible paths of the particle, which
may be formalized as follows. We consider the set, $N$, of paths going from $%
(\mathbf{x}_{1},t_{1})$ to $(\mathbf{x}_{2},t_{2})$, each path specified by
the equation of motion $x^{j}(\nu ;t)$\textbf{, }$\nu \in N,$ $t\in \left[
t_{1},t_{2}\right] ,$ where we assume that x$^{j}$ as a function of $t$, for
fixed $\nu ,$ possesses continuous second derivatives$.$ Our problem is to
find the probability distribution in $N$ corrresponding to that in $\Lambda $%
. Now the condition $\left( \ref{action}\right) $ defines a function, $\nu
=g(\lambda )$, associating a path to every Lagrange function, which may be
got from the variational problem 
\begin{equation}
I\equiv \int_{\Lambda }P\left( \lambda \right) \;\delta \left( \nu -g\left(
\lambda \right) \right) \;dg\left( \lambda \right)
\;\int_{t_{1}}^{t_{2}}L\left( \lambda ;\mathbf{x}\left( \nu ,t\right) 
\mathbf{,\dot{x}}\left( \nu ,t\right) \mathbf{,}t\right) dt=\min ,
\label{action9}
\end{equation}
where $P\left( \lambda \right) $ $\geq 0$ (see $\left( \ref{lagr1}\right) )$
. In fact, any function $g\left( \lambda \right) $ which does not associate
to every $\lambda $ the path giving the minimum value to the action $A\left(
\lambda \right) $ $\left( \ref{action}\right) $ will lead to a value of $I$ $%
\left( \ref{action9}\right) $ which is greater (strictly, not smaller) than
the one given by the function leading to the minimum value of $I.$

In order to devise a practical method to solve the variational problem $%
\left( \ref{action9}\right) $ I suppose that the time integral may be
approximated by a sum over a finite number of times, that is $\varepsilon
,2\varepsilon ,3\varepsilon ,...$ with $\varepsilon $ small enough. For this
we require the position, $\mathbf{y},$ and the velocity, $\mathbf{w},$
corresponding to every path at every one of the discrete times. Thus we
write, instead of $\left( \ref{action9}\right) ,$ the variational condition 
\begin{eqnarray}
J &\equiv &\int_{\Lambda }P\left( \lambda \right) \;\delta \left( \nu
-g\left( \lambda \right) \right) \;dg\left( \lambda \right) \;K=\min ,
\label{act10} \\
K &\equiv &\int_{t_{1}}^{t_{2}}dt\int d^{3}\mathbf{y\;}\int d^{3}\mathbf{w\;}%
\delta (\mathbf{y}-\mathbf{x}\left( \nu ,t\right) )\delta \left( \mathbf{w}-%
\mathbf{\dot{x}}\left( \nu ,t\right) \right) \;L\left( \lambda ;\mathbf{y,w,}%
t\right) ,  \nonumber
\end{eqnarray}
where $\delta \left( {}\right) $ is the three-dimensional Dirac\'{}s delta
and the unknown function to be found is $g\left( \lambda \right) $. Now for
any $g\left( \lambda \right) $ (not necessarily the one providing the
minimum value of $J$ ) we may define a probability distribution, $f_{g}(%
\mathbf{y},\mathbf{w},t)$, in the phase space of positions and velocities
via the integral 
\begin{equation}
f_{g}(\mathbf{y},\mathbf{w},t)=\int_{\Lambda }P\left( \lambda \right)
\;\delta \left( \nu -g\left( \lambda \right) \right) \;dg\left( \lambda
\right) \;\delta (\mathbf{y}-\mathbf{x}\left( \nu ,t\right) )\;\delta \left( 
\mathbf{w}-\mathbf{\dot{x}}\left( \nu ,t\right) \right) .  \label{fg}
\end{equation}
Now our problem is to get, from the variational condition $\left( \ref{act10}%
\right) $, another one involving the phase-space distribution $f_{g}(\mathbf{%
y},\mathbf{w},t),$ rather than the function $g$, if this is possible. The
use of the phase-space distribution leads to a fluid-dynamical picture of
the motion where \textit{the actual particle is replaced by a statistical
ensemble of particles whose phase-space density is }$f(\mathbf{y},\mathbf{w}%
,t)$\textit{. }

A case where the variational problem $\left( \ref{act10}\right) $ may be
easily written in terms of a phase-space distribution corresponds to $L$ not
depending on $\lambda ,$ that is when the Lagrange function is not random. I
shall solve this rather trivial case in the hope that it may provide a clue
for a more general method to be developed in the next section. In this case
inserting $\left( \ref{fg}\right) $ in $\left( \ref{act10}\right) $ leads to
the variational condition 
\begin{equation}
\int_{t_{1}}^{t_{2}}dt\int d^{3}\mathbf{x\;}\int d^{3}\mathbf{w\;}f(\mathbf{x%
},\mathbf{w,}t)\;L(\mathbf{x},\mathbf{w,}t)=\min .  \label{action7}
\end{equation}
Putting the Lagrange function $\left( \ref{lagr1}\right) $ in this
variational problem gives 
\begin{equation}
\int_{t_{1}}^{t_{2}}dt\int d^{3}\mathbf{x\;}\int d^{3}\mathbf{w\;}f(\mathbf{x%
},\mathbf{w,}t)\left[ \frac{1}{2}m\mathbf{w}^{2}-U\left( \mathbf{x}\right)
\right] =\min ,  \label{action8}
\end{equation}
where we ignore the first term of $\left( \ref{lagr1}\right) $ which, being
a constant, is irrelevant in the Lagrange function. Without any loss of
generality we may write 
\[
f(\mathbf{x},\mathbf{w},t)=\rho \left( \mathbf{x,}t\right) h(\mathbf{x},%
\mathbf{w},t), 
\]
with the conditions 
\begin{equation}
\;\int h(\mathbf{x},\mathbf{w},t)d^{3}\mathbf{w=}1,\;\int \mathbf{w}h(%
\mathbf{x},\mathbf{w},t)d^{3}\mathbf{w=v}\left( \mathbf{x,}t\right) ,
\label{fh}
\end{equation}
so that $\rho \left( \mathbf{x,}t\right) $ is the density of particles at $%
\left( \mathbf{x,}t\right) ,$ and $\mathbf{v}\left( \mathbf{x,}t\right) $ is
the mean velocity of those particles present at that space-ime point. Now I
shall make the minimization of $\left( \ref{action8}\right) $ in two steps.
In the first one I will search for the function $h(\mathbf{x},\mathbf{w,}t)$
making the action $\left( \ref{action8}\right) $ a minimum with given $\rho
\left( \mathbf{x,}t\right) $ and $\mathbf{v}\left( \mathbf{x,}t\right) .$ In
the second step I will find the equations of motion of these two functions.

The minimum of $\left( \ref{action8}\right) $ with fixed $\rho $ and $%
\mathbf{v}$ requires, for every $\left( \mathbf{x,}t\right) ,$%
\begin{equation}
\int \mathbf{w}^{2}\mathbf{\;}h(\mathbf{x},\mathbf{w,}t)\;d^{3}\mathbf{w=}%
\min ,\;  \label{vv}
\end{equation}
with the constraints $\left( \ref{fh}\right) ,$ which leads to 
\[
h(\mathbf{x},\mathbf{w,}t)=\delta \left( \mathbf{w}-\mathbf{v}(\mathbf{x}%
,t)\right) , 
\]
$\delta \left( {}\right) $ being the 3-dimensional Dirac\'{}s delta. This
means that all particles present at $\left( \mathbf{x,}t\right) $ possess
the same velocity, that is the phase-space probability becomes 
\begin{equation}
f(\mathbf{x},\mathbf{w},t)=\rho \left( \mathbf{x,}t\right) \delta \left( 
\mathbf{w}-\mathbf{v}(\mathbf{x},t)\right) ,  \label{rov}
\end{equation}
which is a phase-space dependence typical of hydrodynamics. Thus density and
velocity are related by the continuity equation 
\begin{equation}
\frac{\partial \rho }{\partial t}+\mathbf{\nabla j=}0\mathbf{,\;j=}\rho 
\mathbf{v},  \label{cont}
\end{equation}
\textbf{j} being the current density. As a conclusion of the first step, the
form $\left( \ref{rov}\right) $ allows to state the variational condition in
the form 
\begin{equation}
\int_{t_{1}}^{t_{2}}dt\int d^{3}\mathbf{x\;}\rho (\mathbf{x,}t)\left[ \frac{1%
}{2}m\mathbf{v}\left( \mathbf{x,}t\right) ^{2}-U\left( \mathbf{x}\right)
\right] =\min ,  \label{action3}
\end{equation}
with the constraint $\left( \ref{cont}\right) .$

Now we derive the equations of motion of $\rho $ and $\mathbf{v}$ from $%
\left( \ref{action3}\right) ,$ which is straightforward. We introduce the
condition $\left( \ref{cont}\right) $ in the variational problem by means of
the Lagrange multiplier $S(\mathbf{x},t)$ and get 
\begin{equation}
\delta \int_{t_{1}}^{t_{2}}dt\int d^{3}\mathbf{x\;}\left\{ \rho \left[ \frac{%
1}{2}m\mathbf{v}^{2}-U\left( \mathbf{x}\right) \right] +S\left[ \frac{%
\partial \rho }{\partial t}+\mathbf{\nabla }\left( \rho \mathbf{v}\right)
\right] \right\} =0,  \label{action4}
\end{equation}
where the variation of the action must be zero for any independent
variations of $\rho $, $\mathbf{v}$ and $S$ ( the variation of $S$ just
reproduces $\left( \ref{cont}\right) ).$ The function $S$ must fulfil the
condition 
\begin{equation}
S(\mathbf{x}_{1},t_{1})=S(\mathbf{x}_{2},t_{2}),  \label{s}
\end{equation}
relating the initial and the final times and positions. From the variation
of $\mathbf{v}$\textbf{\ }in $\left( \ref{action4}\right) $ we obtain, after
an integration by parts, 
\begin{equation}
\mathbf{v}=\frac{1}{m}\nabla S.  \label{v}
\end{equation}
(The integrated term is zero because $\rho $ vanishes at infinity). If $%
\left( \ref{v}\right) $ is inserted in $\left( \ref{action4}\right) $ we
get, after two appropriate integrations by parts, 
\begin{equation}
\delta \int_{t_{1}}^{t_{2}}dt\int d^{3}\mathbf{x\;}\rho \left[ \frac{1}{2m}%
\left( \mathbf{\nabla }S\right) ^{2}+U\left( \mathbf{x}\right) +\frac{%
\partial S}{\partial t}\right] =0.  \label{action5}
\end{equation}
The time integration by parts requires that the integral 
\[
\int S(\mathbf{x},t)\rho (\mathbf{x},t)d^{3}\mathbf{x\;} 
\]
has the same value at times $t_{1}$ and $t_{2}$, which holds true in view of 
$\left( \ref{s}\right) .$ In the variational problem $\left( \ref{action5}%
\right) $ the variation of $\rho $ leads to the Hamilton-Jacobi equation
whilst the variation of $S$ gives again the continuity eq.$\left( \ref{cont}%
\right) ,$ taking $\left( \ref{v}\right) $ into account, that is 
\begin{equation}
\frac{\partial \rho }{\partial t}+\frac{1}{m}\mathbf{\nabla }\left( \rho 
\mathbf{\nabla }S\right) =0.  \label{cont3}
\end{equation}
These results reproduce standard equations of motion in analytical mechanics
as it should.

\section{Derivation of Schr\"{o}dinger equation}

When $L_{1}\neq 0$, that is the metric is random, it is not obvious that the
problem of the motion may be solved in terms of the two functions $\rho
\left( \mathbf{x,}t\right) $ and $\mathbf{v}\left( \mathbf{x,}t\right) .$
Indeed, the condition $\left( \ref{rov}\right) $ will not be fulfilled in
general. Nevertheless, I shall assume that such a solution of the problem is
still possible provided that we use a \textit{functional} of $\rho \left( 
\mathbf{x,}t\right) $ and $\mathbf{v}\left( \mathbf{x,}t\right) ,$ rather
than a function as in $\left( \ref{action3}\right) .$ In any case the
randomness of the metric will produce randomness in the velocities of the
particles arriving in position $\mathbf{x}$ at time $t$. We shall take into
account that randomness modifying both the continuity eq.$\left( \ref{cont}%
\right) $ and the variational condition $\left( \ref{action3}\right) .$ If $%
\mathbf{w}$ is the velocity of a particle placed in $\mathbf{x}$ at time $t$%
, the position of the particle at time $t+$ $\triangle t$ will be,
neglecting terms of order $\triangle t^{2}$ and higher$,$ 
\begin{equation}
\mathbf{y}=\mathbf{x\ }+\mathbf{w}\triangle t+\triangle \mathbf{x,}
\label{y}
\end{equation}
\textbf{\ \ }$\triangle \mathbf{x}$ being a random displacement. I propose
to treat the displacement $\triangle \mathbf{x}$ as deriving from a white
noise independent of time, position, and initial velocity, thus leading to
the following probability distribution for the (vector) displacement $%
\triangle \mathbf{x}$ 
\begin{equation}
Q\left( \triangle \mathbf{x}\right) =\left( 4\pi D\triangle t\right)
^{-3/2}\exp \left( -\frac{\left| \triangle \mathbf{x}\right| ^{2}}{%
4D\triangle t}\right) .  \label{prob}
\end{equation}
This assumption rests upon the hypothesis that the change of position
induced by the space-time dependence of the coefficients, $g_{\mu \nu }$, of
the metric is more rapid than the change induced by the external forces
(deriving from the potential $U(\mathbf{x})$\ ), which is plausible for a
non-relativistic theory as ours. Actually, if $\left( \ref{prob}\right) $ is
correct, there is a finite (small) probability that the velocity of light is
surpassed, but this fact is not a real problem within the non-relativistic
approximation.

From $\left( \ref{prob}\right) $ it is straightforward to get, from the
density$,$ $\rho \left( \mathbf{x},t\right) ,$ of particles at time $t$ the
density at time $t+\triangle t.$ We get 
\begin{equation}
\rho \left( \mathbf{y},t+\triangle t\right) =\int d^{3}\mathbf{x}\int d^{3}%
\mathbf{w\;}f(\mathbf{x,w\,,}t\mathbf{)}Q(\mathbf{y-x-w}\triangle t),
\label{ro}
\end{equation}
where $f(\mathbf{\mathbf{x,}w\mathbf{,}}t\,\mathbf{)}$ is the phase-space
probability distribution at time $t$. To first order in $\triangle t$ the
integration is straightforward if we use the expansion 
\begin{equation}
f(\mathbf{\mathbf{x,}w\mathbf{,}}t\,)\mathbf{\simeq }f\mathbf{(y\mathbf{%
\mathbf{,}w\mathbf{,}}}t\mathbf{\,\mathbf{)}+}\sum_{j}(x_{j}-y_{j})\frac{%
\partial f}{\partial y_{j}}+\frac{1}{2}\sum_{j}%
\sum_{k}(x_{j}-y_{j})(x_{k}-y_{k})\frac{\partial ^{2}f}{\partial
y_{j}\partial y_{k}}.  \label{fi}
\end{equation}
Inserting $\left( \ref{fi}\right) $ in $\left( \ref{ro}\right) $ and
performing the integrals in $\mathbf{x}$\textbf{\ }and\textbf{\ \thinspace }$%
\mathbf{w}$\textbf{\ }we get, after some algebra, 
\begin{equation}
\frac{\partial \rho }{\partial t}+\mathbf{\nabla j=}0\mathbf{,\;j=}\rho 
\mathbf{v}-D\mathbf{\nabla }\rho ,  \label{cont1}
\end{equation}
where we have taken into account 
\begin{equation}
\int f(\mathbf{y,w\,,}t\mathbf{)d^{3}\mathbf{w}=}\rho \mathbf{\mathbf{(%
\mathbf{\mathbf{x,}}}t\mathbf{\,\mathbf{)}},}\int \mathbf{w\;}f(\mathbf{%
y,w\,,}t\mathbf{)}d^{3}\mathbf{w=}\rho \mathbf{(\mathbf{\mathbf{x,}}}t%
\mathbf{\,\mathbf{)}v(y,}t\mathbf{),}  \label{w}
\end{equation}
(see $\left( \ref{fh}\right) $.) Thus we arrive at the following

\begin{proposition}
The density, $\rho \mathbf{(\mathbf{\mathbf{x,}}}t\mathbf{\,\mathbf{),}}$
and the mean velocity, $\mathbf{v(x,}t\mathbf{),}$ of the particles in the
statistical ensemble representing the actual particle, fulfil de continuity
eq.$\left( \ref{cont1}\right) .$
\end{proposition}

We see that now there is a ``diffusion current'' (last term) in addition to
the ``hydrodynamical current'' (compare with $\left( \ref{cont}\right) ).$
In order to get the substitute for $\left( \ref{action3}\right) $ we
introduce the random velocity $\mathbf{u=w-v(x,}t\mathbf{)},$ so that 
\[
\left\langle \mathbf{w}^{2}\right\rangle _{\mathbf{x},t}=\mathbf{v(x,}t%
\mathbf{)}^{2}+\left\langle \mathbf{u}^{2}\right\rangle _{\mathbf{x},t}+2%
\mathbf{v(x,}t\mathbf{)\cdot }\left\langle \mathbf{u}\right\rangle _{\mathbf{%
x},t}, 
\]
where $\left\langle {}\right\rangle _{\mathbf{x},t}$ means average over
those particles present in $\mathbf{x}$ at time $t$. In this equation the
random-velocity average, $\left\langle \mathbf{u}\right\rangle _{\mathbf{x}%
,t}$ , is the diffusion velocity which, from $\left( \ref{cont1}\right) ,$
should be 
\[
\left\langle \mathbf{u}\right\rangle _{\mathbf{x},t}=-D\rho ^{-1}\nabla \rho
. 
\]
The difficult problem is to calculate the random square mean velocity, $%
\left\langle \mathbf{u}^{2}\right\rangle _{\mathbf{x},t}$, which is not
possible without a detailed knowledge of the probability distribution of
metrics. Thus I will make the most simple assumption, namely that it is a
constant independent of $\mathbf{x}$ and \textit{t}. The lack of a clear
foundation for this hypothesis is certainly a weak point of the present
derivation, which will be studied more carefully elsewhere. In any case,
once the assumption is accepted it is straightforward to arrive at the
following

\begin{proposition}
The mean kinetic energy of the particles (of the statistical ensemble)
present at $(\mathbf{x},t)$ is given by 
\[
T(\mathbf{x},t)=\frac{1}{2}m\left( \mathbf{v}^{2}-2D\rho ^{-1}\mathbf{v\cdot
\nabla }\rho \right) +T_{0}, 
\]
where $T_{0}$ is a constant.
\end{proposition}

The assumption made in $\left( \ref{prob}\right) ,$ that the change of
position induced by the space-time dependence of the metric coefficients is
more rapid than the change induced by the external potential, implies that
the constant $T_{0}$ is rather large so that the kinetic energy is always
positive. As a conclusion the action $\left( \ref{action3}\right) $ should
be replaced by 
\begin{equation}
\int_{t_{1}}^{t_{2}}dt\int d^{3}\mathbf{x\;}L=\min ,\;L\equiv \frac{1}{2}%
m\left( \rho \mathbf{v}^{2}-2D\mathbf{v\cdot \nabla }\rho \right) -\rho U(%
\mathbf{x})  \label{F1}
\end{equation}
where we have ignored the constant $T_{0}$ which is irrelevant in the
variational problem. Now we follow the same steps leading from $\left( \ref
{cont}\right) $ and $\left( \ref{action3}\right) $ to $\left( \ref{action5}%
\right) .$ We introduce the constraint $\left( \ref{cont1}\right) $ in the
variational problem $\left( \ref{F1}\right) $ using a Lagrange parameter S,
which leads to 
\begin{equation}
\delta \int_{t_{1}}^{t_{2}}dt\int d^{3}\mathbf{x\;}\left\{ L+S\left[ \frac{%
\partial \rho }{\partial t}+\mathbf{\nabla }\left( \rho \mathbf{v}-D\mathbf{%
\nabla }\rho \right) \right] \right\} \mathbf{=}0.  \label{F2}
\end{equation}
Hence the variation of \textbf{v} gives, after an integration by parts, the
following relation 
\begin{equation}
\mathbf{v}=D\frac{\mathbf{\nabla }\rho }{\rho }+\frac{\mathbf{\nabla }S}{m}.
\label{vel}
\end{equation}
When this is inserted in $\left( \ref{cont1}\right) $ we get the continuity
equation in the form $\left( \ref{cont3}\right) $. On the other hand when $%
\left( \ref{vel}\right) $ is inserted in $\left( \ref{F2}\right) $ we obtain 
\begin{equation}
\delta \int_{t_{1}}^{t_{2}}dt\int d^{3}\mathbf{x\;}\left\{ S\frac{\partial
\rho }{\partial t}-\rho \left[ \frac{1}{2m}\left( \mathbf{\nabla }S\right)
^{2}+\frac{1}{2}mD^{2}\left( \frac{\mathbf{\nabla }\rho }{\rho }\right)
^{2}+U\left( \mathbf{x}\right) \right] \right\} =0,  \label{action6}
\end{equation}
where we have ignored two terms whose sum equals the divergence of the
vector field $DS\mathbf{\nabla }\rho .$ We must assume that this vector
field vanishes at infinity whence the integral of its divergence is zero.

It is remarkable that, although we have started from two expressions, $%
\left( \ref{cont1}\right) $ and $\left( \ref{F1}\right) ,$ neither of which
is invariant under the reversal of time, the Lagrange density in $\left( \ref
{action6}\right) $ \textit{is time-reversal invariant} in the sense that the
operation ($t\rightarrow -t)$ is equivalent to just changing the sign of the
auxiliary function \textit{S}, which does not change the physics. In $\left( 
\ref{action6}\right) $ it is easy to see that the variation of $\rho $ leads
to 
\begin{equation}
\frac{1}{2m}\left( \mathbf{\nabla }S\right) ^{2}+U\left( \mathbf{x}\right) +%
\frac{\partial S}{\partial t}-2mD^{2}\frac{\nabla ^{2}\sqrt{\rho }}{\sqrt{%
\rho }}=0,  \label{sch1}
\end{equation}
where we have assumed that the space-time integral of $\partial \left( S\rho
\right) /\partial t$ vanishes, a hypothesis already made in the previous
section (see below eq.$\left( \ref{action5}\right) )$. As is well known the
continuity eq.$\left( \ref{cont3}\right) $ and the dynamical eq.$\left( \ref
{sch1}\right) $ may be obtained by separating the real and imaginary parts
in the Schr\"{o}dinger equation 
\begin{equation}
i\hbar \frac{\partial \Psi }{\partial t}=-\frac{\hbar ^{2}}{2m}\nabla
^{2}\Psi +U\Psi .  \label{sch}
\end{equation}
provided that we identify 
\begin{equation}
\hbar \equiv 2mD,\;\Psi \equiv \sqrt{\rho }\exp \left( \frac{iS}{\hbar }%
\right) .  \label{h}
\end{equation}

\section{Discussion}

The derivation of Schr\"{o}dinger equation given in this paper might be
considered a step in the direction of Einstein's expectations, as commented
in the introduction. Alternatively it may be seen as just a new derivation
from\textit{\ formal }assumptions (the two propositions of the previous
section) devoid of any deep physical meaning. I cannot argue too strongly in
favour of the former possibility, and this is why I include the word
``heuristic'' in the title of the paper.

In our derivation the ``wave-function'' $\Psi $ is just a mathematical
function, with range in the complex numbers, whose modulus gives information
about the probability density of the particle position and whose phase is
related to the particle velocity via eq.$\left( \ref{vel}\right) $. In
comparison with the standard Hamilton-Jacobi equation, see $\left( \ref
{action5}\right) ,$ eq.$\left( \ref{sch1}\right) $ contains a term deriving
from the fact that the space-time metric differs from the standard one of
Minkowski space. Thus the metric plays the role of the ``guiding wave'' in
the de Broglie-Bohm theory\cite{Holland}. It is not contrary to the
intuition that this ``wave'' may modify the scattering cross sections,
making them different from those derived from classical mechanics. Also it
is easy to understand the existence of stationary states of electrons in
atoms, as a balance between the attraction by the nucleus and the diffusion
caused by the random metric. The picture is here similar to that provided by
stochastic electrodynamics\cite{Luis} with the gravitational field (the
non-Minkowskian metric) substituted for the electromagnetic radiation. More
difficult is to understand intuitively how the non-Minkowskian metric may
give rise to the observed sharp spectral lines of atoms or to the
interference fringes in two-slit experiments. If the latter two effects
derive from the metric, the mechanism is not clear from the derivation here
presented.

A question which arises is whether the assumption of a random metric is
really necessary for the two hypotheses introduced in the previous section.
Indeed, derivations of Schr\"{o}dinger equation from the hypothesis of a
random motion, not involving the space-time metric, have been recurrently
proposed during the last 50 years or more\cite{Nelson},\cite{Luis}. In my
view there are three reasons why the randomness of the metric is relevant in
the derivation given in the present paper. The first one is the fundamental
character of the space-time background, as emphasized in the introduction.
The second reason is the existence of a diffusion in coordinate space rather
than in velocity space, as would be more natural in any approach starting
from Newtonian mechanics. Also, if the origin of the random motion is
gravitational, dissipative effects may be negligible. The third reason is
that metric randomness makes plausible a formulation starting from a
variational condition, due to the fundamental role which geodesics play in
general relativity. Nevertheless the fact that Planck\'{}s $\hbar $ is a
universal constant barely follows from the existence of random metrics.
Indeed within general relativity (or Newtonian gravity) the motions (of
small particles) are independent of the mass, which suggests that the
diffusion constant $D$ (see $\left( \ref{h}\right) )$ , rather than $\hbar ,$
should be universal. I have no clear response to this objection. I may
mention only the fact that, when we consider the diffusion of a system
consisting of $n$ particles of equal mass, $m$, it may be shown that the
diffusion constant of the center of mass of the system is $D/n$ if the
particles diffuse independently. In this case the product, $mD$, of the mass
times the diffusion parameter is the same for the individual particles and
for the center of mass of the system, independently of $n$.

An interesting question is whether the derived equations ($\left( \ref{cont3}%
\right) $ and $\left( \ref{sch1}\right) )$ are equivalent to Schr\"{o}dinger
equation. The answer is in the negative. The reason is that the derivation
of the previous section implies that the function $S\left( \mathbf{x}%
,t\right) $ must be single-valued whilst in Schr\"{o}dinger theory the
wavefunction $\Psi \left( \mathbf{x},t\right) $ is single-valued, which just
requires that $S\left( \mathbf{x},t\right) $ changes by an integer multiple
of $2\pi \hbar $ along any closed line. Actually both assumptions are
equivalent, by continuity, if the region of definition of $S$ (or $\Psi )$
is simply connected. However it is currently assumed that there are
experimental situations where this is not the case. In particular the
popular two-slit experiments, which have already been performed with
electrons, neutrons, atoms and even molecules, are currently analyzed
assuming the existence of a region forbidden to the particles, namely the
one occupied by the screen with the slits, so that the region allowed to the
particles is not simply connected. However it might be possible to study
these experiments replacing the screen by a high, but finite, repulsive
potential so that the region is simply connected. Another case where the
region of interest is not simply connected corresponds to some excited
states of atoms. In this case the wavefunction may become singular at the
origin (the position of the nucleus). For instance the wavefunction of the
states with quantum numbers $l=m\neq 0$ contains the factor $exp(il\phi )$
and the function $S$ changes by $2\pi l\hbar $ in a rotation by $2\pi .$ A
related problem is the existence of nodal surfaces in some solutions of
Schr\"{o}dinger equation, that is surfaces where $\rho =0$ but $\mathbf{v}%
\neq 0. $ These solutions are unphysical in our approach. In summary every
physical solution of $\left( \ref{cont3}\right) $ and $\left( \ref{sch1}%
\right) $ \textit{is} a solution of Schr\"{o}dinger equation $\left( \ref
{sch}\right) , $ but \textit{there are solutions of the latter which are not
solutions of the former}. This seems to imply that the formalism here
developed cannot agree with the empirical evidence. Nevertheless I question
the current wisdom that \textit{all} solutions of the Schr\"{o}dinger
equation are really necessary for the interpretation of the experiments, but
this point will not be analyzed further here.

In any case the formulation here presented allows for an interpretation of
Schr\"{o}dinger equation in terms of trajectories, which may be useful in
some applications. In this sense the formulation is similar in spirit to
Bohmian mechanics\cite{Holland}. However our trajectories present a random
element and so the picture achieved is actually more similar to stochastic
mechanics. But stochastic mechanics is currently understood as fully
equivalent to Schr\"{o}dinger theory, which leads to counterintuitive
behaviour like the existence of nodal surfaces mentioned above. For this
reason it is usually considered as just a formal approach to quantum
mechanics rather than a different physical theory\cite{Nelson}.

The formalism here presented might be extended to many-particle systems by
replacing the three dimensional space by the $3N$ dimensional configuration
space of $N$ particles. However all particles will move in the same
space-time metric, which would induce correlations in the motion, in
addition to those derived from the possible inter-particle forces. Thus the
generalization is not trivial. Incidentally, I guess that the said
correlations might be related to Bose statistics, but this point will not be
discussed further here.

In summary, the formalism here developed allows an interpretation of
Schr\'{o}dinger equation in terms of particle trajectories, which may have
some interest. However there are great difficulties to take it as a physical
theory underlying quantum mechanics.

\end{document}